%% file: main.tex
\documentclass[10pt,journal,romanappendices]{IEEEtran}
\input{my_macros.tex}

\usepackage{algpseudocode}
\usepackage{verbatim}
\usepackage{multirow}
\usepackage{url}
\usepackage{makecell}
\usepackage{float}
\usepackage{comment}
\usepackage{enumitem}
\usepackage{subfigure, caption}
\usepackage{subcaption}
\usepackage{multirow}
\usepackage{bbding}
\usepackage{soul}
\newcolumntype{C}[1]{>{\centering\arraybackslash}p{#1}}

\begin{document}

\title{Topology-Aware Coordination for Multi-Functional Low-Altitude Wireless Networks}

\author{Jiajun He,~\IEEEmembership{Member, IEEE,}~Han Yu,~\IEEEmembership{Member, IEEE},~Yiran Guo,~Xinping~Yi,~\IEEEmembership{Senior Member, IEEE},\\
~Fan~Liu,~\IEEEmembership{Senior Member, IEEE},~Hing Cheung So,~\IEEEmembership{Fellow, IEEE},
Hien Quoc Ngo,~\IEEEmembership{Fellow,~IEEE,}\\ 
~Michail Matthaiou,~\IEEEmembership{Fellow,~IEEE},~and~Giuseppe Caire,~\IEEEmembership{Fellow, IEEE} 
\thanks{J. He, H. Ngo, M. Matthaiou are with the Centre for Wireless Innovation (CWI), Queen’s University Belfast, BT3 9DT Belfast, U.K. (email: \{j.he, hien.ngo, m.matthaiou\}@qub.ac.uk).}
\thanks{H. Yu and G. Caire are with the Faculty of Electrical Engineering and Computer Science, Technical University of Berlin, Germany (email:han.yu.1@campus.tu-berlin.de, caire@tu-berlin.de).}
\thanks{Y. Guo is with the School of Electronic and Information Engineering, Beijing Jiaotong University, Beijing, China. (e-mail: yiranguo@bjtu.edu.cn)}
\thanks{X. Yi and F. Liu are with the National Mobile Communications Research Laboratory, Southeast University, Nanjing 210096, China (e-mail: \{xyi, fan.liu\}@seu.edu.cn).}
\thanks{H. C. So is with the Department of Electrical Engineering, City University of Hong Kong, Hong Kong SAR, China (email: hcso@cityu.edu.hk).}
\thanks{\textit{(Corresponding Author: Han Yu.)}}
}

\markboth{}%
{Shell \MakeLowercase{\textit{et al.}}: Bare Demo of IEEEtran.cls for IEEE Journals}

\maketitle
\vspace{-1.5em}
\begin{abstract}
Low-altitude wireless networks (LAWNs) are expected to consist of multi-tier, heterogeneous terrestrial and non-terrestrial devices, where effective coordination is essential to fully unlock the complementary capabilities of diverse systems from different vendors. To address this issue, we propose a novel  multi-functional coordination framework that enables seamless cooperation within the LAWN while supporting efficient execution of diverse network functions. In the proposed architecture, each device or infrastructure element is assigned to a specific functional role, namely, edge mobile terminal (E-MT), distributed MT (D-MT), or computing center. E-MTs are equipped with lightweight, independent signal processing and computing capabilities, while D-MTs and the computing center handle regional and global coordination, respectively. To enhance the overall network efficiency, we model the LAWN as a sparse graph, where nodes represent network nodes and edges are defined according to a set of controllable connection rules. This topology-aware (TA) representation allows for efficiently solving various coordination tasks across the network. Numerical results show that the proposed TA coordination framework outperforms baseline approaches that lack topological insights, achieving higher efficiency in multi-task coordination. Finally, we discuss key technical challenges and outline potential solutions for future deployment.
\end{abstract}


\IEEEpeerreviewmaketitle

\section{Introduction}

The sixth-generation (6G) and beyond networks are anticipated not only to support ultra-high data rates but also to deliver various services in regions where terrestrial infrastructure is not available \cite{11358925}. The rapid advancement of airborne communication platforms, such as unmanned aerial vehicles (UAVs) and electric vertical takeoff and landing (eVTOL) aircrafts, is fostering the development of a low-altitude digital service ecosystem, referred to as the low-altitude economy. Due to their high deployment flexibility and favorable dynamic line-of-sight (LoS) characteristics, airborne platforms are able to overcome the limitations of terrestrial networks and to extend wireless services to underserved or unreachable areas. In a low-altitude wireless network (LAWN), various airborne terminals may exhibit different capabilities and functional roles, depending on their design and application scenarios \cite{Energy_Harvesting, LAWN_Survey}. In general, they can be categorized into multiple groups based on their designated functions, such as:

\begin{enumerate}
    \item \textit{Environmental Sensing}: In LAWNs, UAVs can be equipped with specialized sensors, such as cameras, light detection and ranging (LiDAR), and millimeter-wave (mmWave) radar, to perform high-resolution environmental mapping, object detection, and target tracking. The sensing capabilities of LAWNs are essential for enabling real-time situational awareness and accurate target localization.

    \item \textit{Personal Network and Relay}: Airborne platforms in LAWNs can function as aerial base stations (BSs) or mobile relays to extend coverage and enhance capacity in areas with limited or no terrestrial infrastructure \cite{7572068}. By dynamically adjusting their positions, these UAVs are able to mitigate coverage gaps and improve signal quality in dynamic or high-demand environments. In addition, UAVs can act as personalized cells (pCells), forming dedicated, user-centric links that follow and serve individual users, thereby enabling ultra-reliable and adaptive communication.

    \item \textit{Wireless Power Transfer (WPT)}: Equipped with radio-frequency (RF) transmission technologies, UAVs can deliver wireless power to ground devices or nearby airborne terminals. This capability is particularly advantageous for lightweight UAVs with limited battery capacity acting as energy receivers, since the external power supporter is essential to sustain extended operational endurance. 
\end{enumerate}

Nevertheless, the complex network architecture and highly dynamic environment of LAWNs present significant challenges in supporting multiple functions simultaneously, while the strong coupling among multiple objectives, combined with the time-varying behavior of UAV platforms, further complicates the coordination process. In addition, the multi-objective optimization problems in LAWNs are non-convex and computationally intensive, making real-time solutions difficult to achieve. Motivated by this challenge, a weighted objective formulation was introduced in \cite{MO_SC2} to balance communication performance and sensing accuracy in LAWNs and is solved via alternating optimization to coordinate resource allocation and UAV deployment. Sun $et$ $al.$ \cite{MO_Edge} studied the joint optimization of latency, energy consumption, and task offloading efficiency in UAV-assisted mobile edge computing, where the resulting complex optimization problem was addressed using a block alternating descent (BAD) method. Considering the significant role of WPT in 6G and beyond networks, \cite{li2024iscpt} introduced an integrated sensing, communication, and power transfer (ISCPT) framework that unifies integrated sensing and communication (ISAC) with simultaneous wireless information and power transfer (SWIPT), aiming to increase both energy and spectral efficiency. Building upon this, \cite{zhou2024iscpt} presented a unified framework for ISCPT, where the joint optimization is formulated as a non-convex problem and efficiently solved using semidefinite relaxation (SDR). Moreover, Chu $et$ $ al.$ \cite{chu2025revolutionizing} considered the ISCPT problem, which aims to maximize energy harvesting using particle swarm optimization, while simultaneously satisfying the strict communication quality-of-service (QoS) requirement and sensing accuracy in terms of the Cramér-Rao bound (CRB). Nevertheless, the existing studies mainly rely on alternating optimizations or heuristic solutions, which often suffer from high computational complexity.

{To our knowledge, most existing studies on wireless communication primarily focus on signal-level modeling and often overlook the structural characteristics of the network, typically assuming fully connected systems. However, in realistic scenarios, connectivity among nodes is highly selective and heterogeneous, as the channel strength is strongly influenced by factors, such as blockage, distance, and operating frequency bands.
Such selective connectivity can be naturally abstracted through a topological structure, which models the network as a structured, sparse graph capturing feature-dependent relationships among nodes. 
By exploiting the inherent topology of the terrestrial network, various functions, such as user scheduling and pilot assignment \cite{TA_terrestrial}, can be efficiently performed without relying on instantaneous channel state information (CSI) and frequent information exchange. This topology-aware (TA) approach enables enhanced system performance with reduced signaling overhead and improved scalability. }

In LAWNs, the network naturally forms a large-scale topological structure, where each device can be abstracted as a node with distinct functional characteristics, while the relationships between nodes are represented by edges. As a result, the multi-functional LAWN can be represented on a graph, enabling the corresponding optimization problems to be addressed through equivalent graph-based formulations. Given the joint constraints involving multiple objectives, the proposed framework can provide physical interpretations for each objective, enhancing the generalizability and practical applicability of the LAWN. By dynamically updating node attributes and edge weights, this framework enables real-time adaptation of link conditions and information transmission states among numerous low-altitude devices, thereby addressing the core challenges in LAWNs.

In this paper, we investigate a LAWN framework comprising both terrestrial and non-terrestrial mobile terminals (MTs), where each MT assumes a distinct role within the network, including edge MTs (E-MTs) and distributed MTs (D-MTs).
Computationally intensive tasks are offloaded to terrestrial gateways or satellite computing centers for coordination processing when necessary. The guiding principle of the TA-based framework is a low-overhead and efficient coordination strategy that minimizes reliance on exhaustive real-time data exchange by exploiting the relatively stable and predictable topological patterns of LAWNs. This design is particularly beneficial in scenarios requiring frequent coordination, such as UAV trajectory planning and dynamic resource allocation. Our approach allows each MT to retain local processing autonomy while enabling efficient coordination between E-MTs and D-MTs. Moreover, the framework preserves the structural advantages of a hierarchical architecture while significantly reducing signaling overhead, thereby achieving a practical balance between global optimization and distributed decision-making to enhance overall system performance and scalability in dynamic low-altitude environments.

The remainder of this article is organized as follows: Section \ref{Section_System_Model} introduces the basic concepts and inherent coordination mechanisms of the proposed TA-based LAWN framework, along with simulation results that demonstrate its advantages in Section \ref{Section_Case_Studies}. Section \ref{Section_Challenges} discusses challenges and potential solutions. Finally, Section \ref{Section_Conclusion} concludes the article.


\section{Multi-Functional LAWN Framework}
\label{Section_System_Model}
 
\begin{figure*}[!t]
\centering
\includegraphics[width=6in]{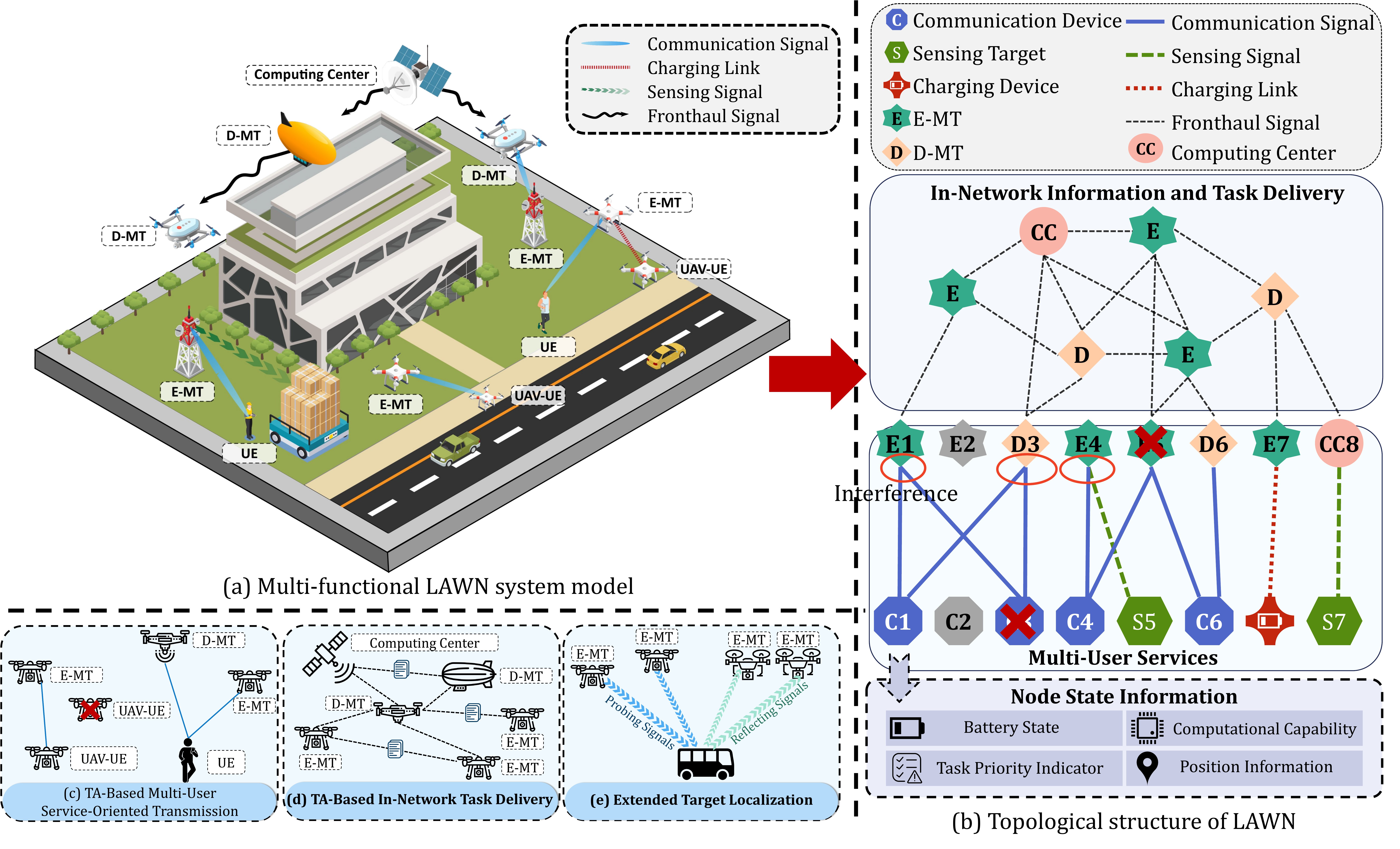}
\caption{Overview of multi-functional LAWN and its graph representation.}
\label{fig:SystemModel}
\vspace{-0.7cm}
\end{figure*}



\subsection{Multi-Tier Multi-Functional LAWN Network}

To support an efficient LAWN operation, a low-overhead TA-based coordination framework is introduced, which operates within a hierarchical architecture that assigns distinct roles to different types of MTs. Specifically, the E-MTs are responsible for performing diverse functions, while D-MTs handle coordination tasks within a certain region. The computationally intensive workloads from D-MTs may be offloaded to a terrestrial gateway or a satellite computing center for coordination processing. This framework offers flexibility in dynamically reconfiguring the MT roles based on evolving task requirements. We begin by presenting the key components of this framework, as illustrated in Fig. \ref{fig:SystemModel}(a).

\begin{itemize}[leftmargin=10pt]
    \item \textit{E-MT}: In the TA-based framework, an E-MT is assigned to perform specific functions based on task requirements and prevailing system conditions. An E-MT can be either a mobile airborne terminal, such as a UAV, or a terrestrial access point (AP), depending on the deployment scenario. Each E-MT is equipped with a local processing unit capable of autonomously handling lightweight tasks, while computationally demanding tasks are offloaded to upper-layer units for further processing. It is worth noting that E-MTs can collaborate with neighboring E-MTs via sidelink communication, such as device-to-device (D2D) links, to enable peer-level support.
    
    \item \textit{D-MT}: The D-MT functions as a regional controller, overseeing clusters of E-MTs within a designated airspace. The D-MT is typically equipped with moderate onboard computing resources and enhanced communication capabilities, while its primary responsibilities include:
    
    \begin{enumerate}
    
    \item \textit{Topology-Aware Coordination}: Managing and scheduling E-MT operations based on network topology to optimize task distribution, interference management, and spatial coverage.
    
    \item \textit{Resource Allocation and Task Offloading}: Coordinating multiple functions within E-MT clusters to avoid collisions and maximize operational efficiency. In addition, the D-MT receives and processes tasks offloaded from E-MTs that exceed their local processing capacity.
    
    \item \textit{Interfacing with Computing Centers}: Forwarding high-complexity or delay-tolerant tasks to the computing center and receiving global optimization directives to support large-scale coordination.
    \end{enumerate}
    
    By aggregating local network state information, D-MTs serve as a middle-tier in the LAWN architecture, enhancing the scalability and robustness of the network.

    \item \textit{Computing Centers}: The computing center, whether a terrestrial gateway or a satellite in low earth orbit (LEO) or geostationary orbit (GEO), is responsible for global coordination and long-term optimization across the LAWN. The satellite center is activated as a backup when terrestrial communication links are disrupted, ensuring continued coordination and network functionality. Through periodic assessments of network conditions, such as throughput and sensing accuracy, it enables global resource optimization and facilitates seamless integration between low-altitude platforms and terrestrial infrastructure.
\end{itemize}

\vspace{-0.5cm}

\subsection{Topology Aware-LWAN Framework}
\label{sec:TAframework}
This subsection demonstrates how the LAWN operational problem can be reformulated as a TA problem. As illustrated in Fig. \ref{fig:SystemModel}(b), all network nodes are abstracted as either transmitter nodes or user nodes, without explicitly distinguishing between non-terrestrial and terrestrial units. The connections between these nodes are represented by edges, which capture the relationships and interactions among them. The definitions and key characteristics of the node and edge are as follows:
\begin{itemize}[leftmargin=10pt]
    \item \textit{Nodes}: In the TA-based framework, all participating devices are abstracted as nodes characterized by their unique features, while the relationships among them are represented by edges. In general, node features are embedded with task priority indicators. A node may be assigned multiple concurrent tasks, potentially leading to task conflicts. In such cases, a single task is selected as the priority active, while the node status is updated accordingly to its current operational.
    
    \textbf{Example}: Battery state is a critical feature for the UAV. When the remaining energy falls below a certain level, the UAV is classified as a charging user. As the battery level continues to decrease, its task priority correspondingly increases, prompting the network to allocate charging resources preferentially to that node. In addition, computational capability is another important node feature. If the workload assigned to a node exceeds its processing capacity, part of the task is offloaded or reassigned to neighboring nodes. To summarize, incorporating these node features enables more efficient task scheduling and resource coordination.
    
    \item \textit{Edges}: Edges characterize the relationship among nodes, with their definitions adaptable to the particular objectives and constraints of the optimization problem.
    
    \textbf{Example}: In LAWNs, the connectivity between nodes is typically determined by the geometric distance or channel conditions. An edge is established between two nodes if their channel gain exceeds a predefined threshold, or if their distance and blockage conditions satisfy the minimum connectivity criteria. The existence of such an edge implies that the two nodes can support reliable communication. In addition, an edge can characterize interference relationships. For instance, an edge may be created between two users if they are expected to cause significant interference. This dual interpretation allows the TA model to capture both connectivity and interference dynamics in the network.
    
    \item \textit{Rules}: To support diverse network functionalities, the graph abstraction of the LAWN is augmented with a set of control rules that govern the node behavior. These rules explicitly define which node and edge features are controllable, thereby identifying the decision variables for the subsequent optimization process. By distinguishing between controllable parameters (e.g., transmit power, beamforming direction, or task allocation) and network attributes (e.g., physical location or hardware limitations), the framework enables a systematic and tractable approach to topology-driven optimization.
    
    \textbf{Example}: The basic rules of the TA-based framework are governed by the binary activation states of the nodes, which serve as a key controllable parameter in the LAWN design and optimization. The existence of an edge is determined by the activation states of its associated nodes, while an edge is established only when both endpoint nodes are active. 
    Power allocation can also be defined as another controllable parameter. Although it does not influence the existence of edges, it can affect the edge weight, thereby influencing the optimization and enabling effective solution strategies.
\end{itemize}

It is worth noting that the definitions presented are problem-dependent and can be flexibly adapted by modifying node and edge attributes to suit specific application requirements.

\vspace{-0.3cm}

\subsection{Superiority of TA-based Framework}

In this subsection, we first present the interference avoidance as an illustrative example to clearly show the proposed TA framework and demonstrate its application to a simple LAWN scenario. More complex examples are provided in Section \ref{Section_Case_Studies}. We then highlight the key advantages of the proposed TA-based framework, with particular emphasis on its natural compatibility with graph neural networks (GNNs) and its strong generalization capability and flexibility in supporting diverse network functionalities and dynamic topological conditions.

{ \subsubsection{Illustrative Example} In the interference avoidance problem, our objective is to select users in a manner that minimizes interference. When formulated purely from a communication perspective, this problem can be difficult to interpret. However, it becomes more intuitive when viewed through the proposed framework. Specifically, the communication system is first represented as a bipartite graph, where transmitters and receivers are placed on opposite sides of the graph. An edge indicates that the channel strength between two nodes is sufficiently high. Overlaps on the transmitter side of the bipartite graph represent potential interference among users. These overlaps can be managed by activating or deactivating selected transmitters and receivers. For example, as illustrated in Fig.~\ref{fig:SystemModel}(b), overlaps occur at nodes E1, D3, and E5. By deactivating nodes E5 and C3, the interference among users can be effectively mitigated. In this context, the interference avoidance problem is naturally characterized as the elimination of overlaps in the bipartite graph, while Fig.~\ref{fig:SystemModel}(b) also illustrates a feasible solution.}


\subsubsection{Connection with GNN}

The TA-based framework naturally aligns with GNNs for several key reasons \cite{11232498}. First, this framework abstracts the network as a graph, where nodes and edges are characterized by features and activation states that directly correspond to the input structure of GNNs. Second, the multi-functional objectives of the LAWN can be embedded into the GNN training process via well-defined loss functions. Moreover, our framework is compatible with reinforcement learning paradigms \cite{11237077}, where the physical interpretation of multi-functional network operations can be leveraged to guide the design of reward functions. This integration enables more effective policy learning and decision-making across dynamic and heterogeneous network environments.

\subsubsection{Generalizability and Flexibility}

In both traditional model-based and neural network-based optimizations, generalizability and flexibility are critical. For neural network methods, changes in the system environment, such as variations in scattering or blockage conditions, often require re-training the network, since the pre-trained model is tightly coupled to a specific operating scenario. In contrast, our framework enables multi-functional problems to be represented and processed within a graph-based abstraction. The neural network is trained to operate on the graph structure, where the loss function is determined by node and edge attributes rather than a fixed system configuration. As the system evolves, system variations are captured by changes in network topology instead of the problem formulation, which removes the need for retraining the model.

Moreover, for most traditional optimization-based methods, variations in system objectives or constraints generally require either a redesign or a reformulation of the optimization problem. In contrast, the devised approach focuses on modeling relationships among nodes, such as resource allocation, interference coupling, and degrees-of-freedom (DoFs), which are fundamental elements shared by a wide range of objectives in heterogeneous networks. Consequently, diverse objectives can be supported by properly incorporating these elements into the framework, while keeping the core methodology unchanged under varying optimization goals.

\vspace{-0.2cm}

\section{Case Studies and Applications}
\label{Section_Case_Studies}

In this section, we present three representative use cases as shown in Figs. \ref{fig:SystemModel}(c) to \ref{fig:SystemModel}(e) to illustrate the functionality and effectiveness of the designed TA-based LAWN framework, including: 1) TA-based multi-user service-oriented transmission, 2) TA-based in-network task delivery, and 3) extended-target localization. These scenarios demonstrate the flexibility of our framework in supporting diverse operation scenarios.

\begin{figure*}
  \begin{minipage}[b]{2\columnwidth}
  \centering
  \subfigure[]{
  {\includegraphics[width=0.3\linewidth]{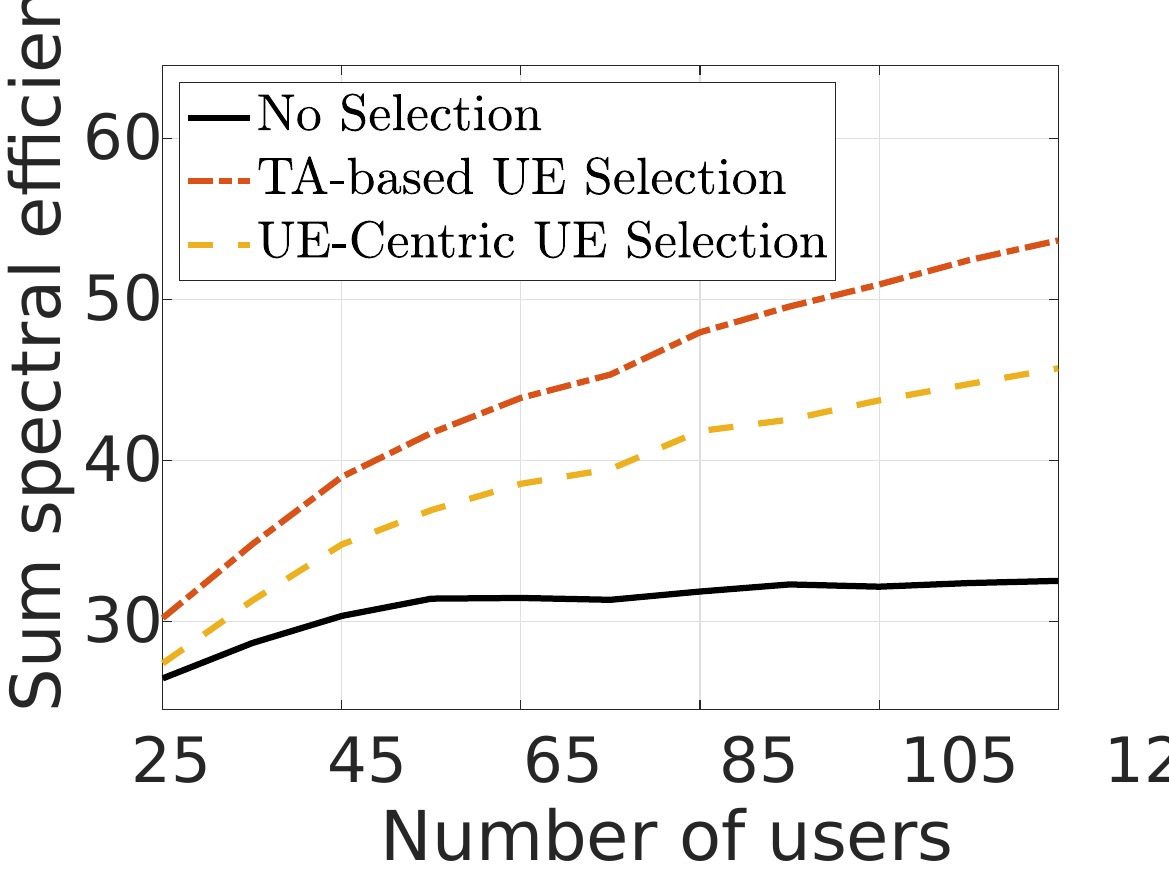}}
  \label{fig-RSS hybrid corner}
  }
  \subfigure[]{
  {\includegraphics[width=0.3\linewidth]{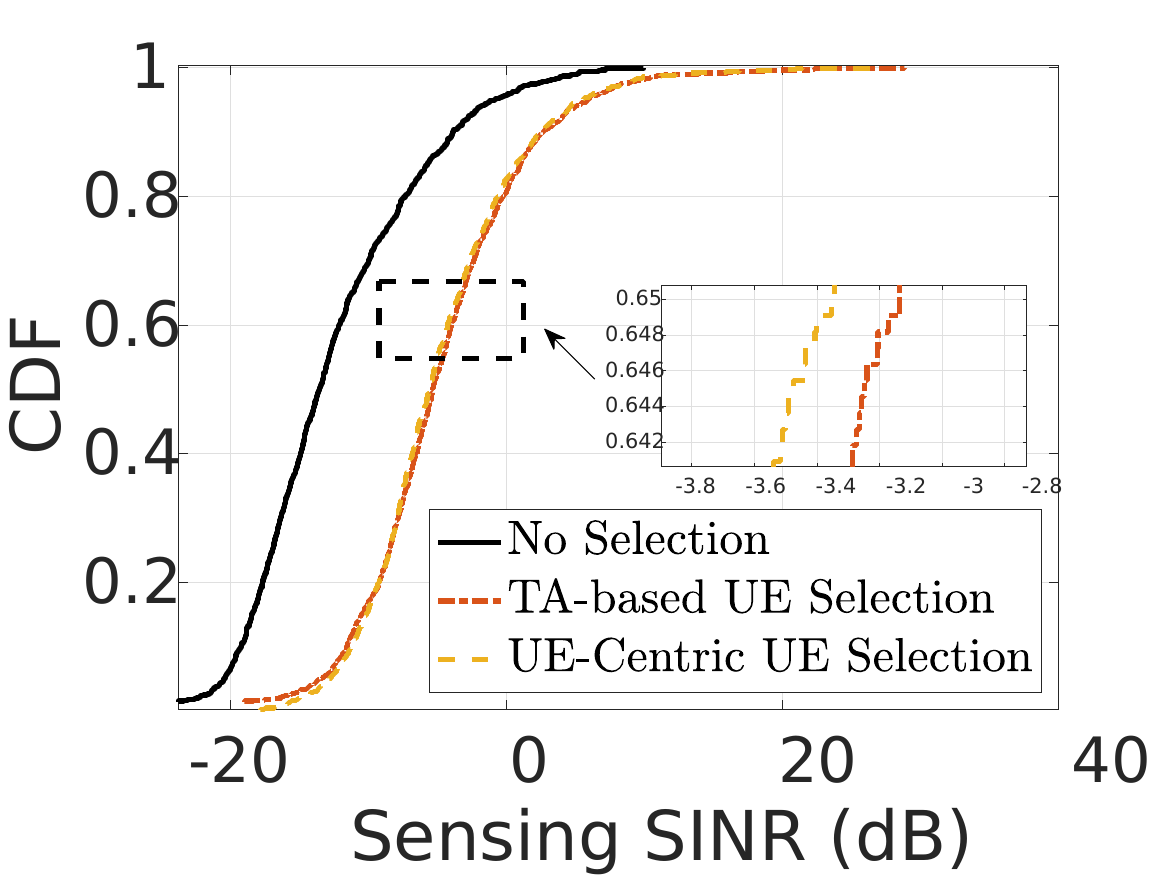}}
  \label{fig-loc meeting room}
  }
  \subfigure[]{
  {\includegraphics[width=0.3\linewidth]{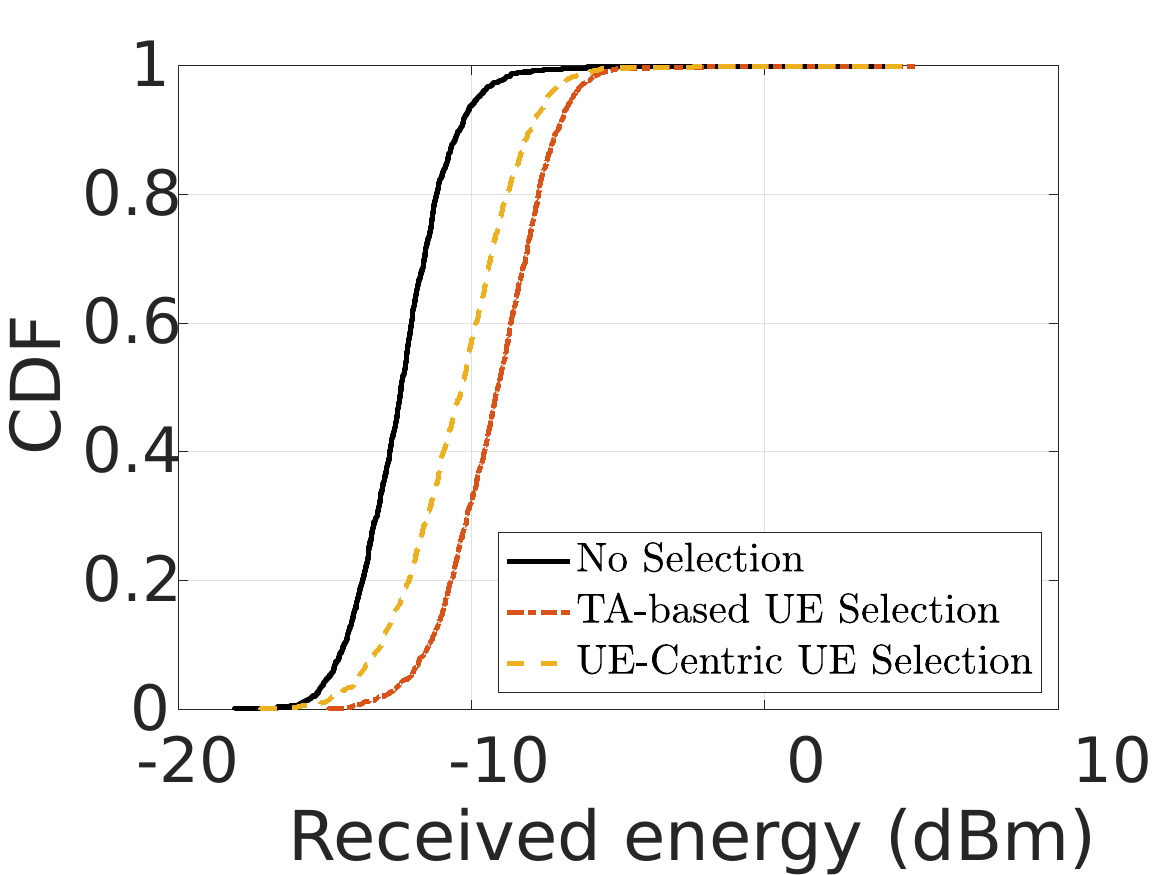}}
  \label{fig-sep distance}
  }
  \caption{The figures depict the performance of a multi-user, multi-function problem: Fig. \ref{fig-RSS hybrid corner} is the communication performance through sum spectral efficiency; Fig. \ref{fig-loc meeting room} is the sensing SINR of the target, while Fig. \ref{fig-sep distance} is the total received energy by all charging users. }
  \label{fig:ISCPT}
  \end{minipage}
  \vspace{-0.2cm}
\end{figure*}


\subsection{Simulation Setups}
\label{Subsection_Simulation_Setups}

In our simulation, UAVs and terrestrial APs are randomly deployed within a three-dimensional space with dimensions $2$ km $\times$ $2$ km $\times$ $50$ m. Specifically, $64$ UAVs and $64$ terrestrial APs as E-MTs are uniformly distributed over the $2$ km × $2$ km horizontal plane, and the altitudes of both E-MT UAVs and user UAVs randomly assigned between $10$ m and $50$ m. Each E-MT is equipped with a $2 \times 2$ uniform planar array (UPA) to enable directional transmission. This scenario involves multiple users, with $80\%$ comprising UAV users and the remaining $20\%$ being terrestrial users. We consider four charging users and one sensing target. In addition, we focus on the $2.6$ GHz frequency band within the LAWN framework, while the free-space path loss \cite{7572068} is employed to evaluate channel strength, which is then used to construct the topological structure of the network.

\vspace{-0.4cm}

\subsection{TA-Based Multi-User Service-Oriented Transmission}

\begin{figure}
    \centering
\includegraphics[width=0.65\linewidth]{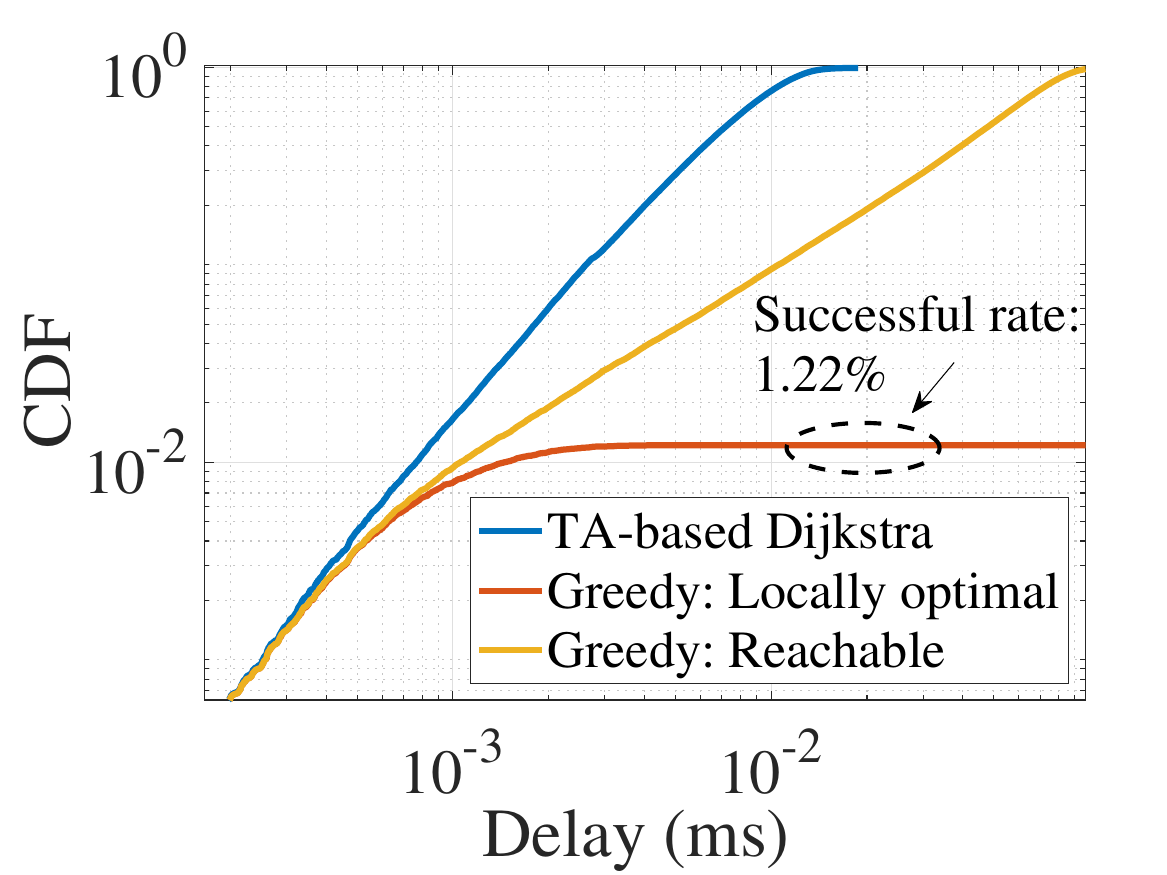}
    \caption{Delay comparison between the proposed TA-based scheme and greedy-based locally optimal and reachable schemes during in-network task delivery phase.}
    \label{fig:delay}
    \vspace{-0.6cm}
\end{figure}

This subsection studies a user selection problem, as shown in Fig. \ref{fig:SystemModel}(c) in an overloaded LAWN that simultaneously supports communication, target sensing, and WPT, while the typical non-selection and user-centric selection schemes are served as the benchmarks to evaluate the effectiveness of the proposed TA-based framework. Note that for the users, only the activation states of communication users are selectable, while the sensing target and charging users are always active.

In the considered multi-functional LAWN, efficient communication requires a high received signal strength (RSS) as well as a low-interference operating environment. From a sensing perspective, accuracy can be significantly degraded by the communication service, as communication signals introduce additional interference that impairs sensing performance. In addition, during the downlink phase, the WPT signal does not interfere with other services; however, it consumes transmit power and must be accounted for in the power allocation constraints. Figure \ref{fig:ISCPT} illustrates the performance of spectral efficiency, sensing SINR, and total received energy at the WPT users. The `No Selection' lines indicate that the system is overloaded, making the selection essential. Since the user-centric solution fails to account for these interdependent factors, it is observed that the TA-based scheme consistently outperforms the user-centric strategy by more effectively accommodating all three functional objectives.

\vspace{-0.2cm}

\subsection{TA-Based In-Network Task Delivery}

The second case study (Fig. \ref{fig:SystemModel}(d)) focuses on distributing control instructions and coordinating service-triggering tasks across the LAWN. In this context, only connections with RSS exceeding a given threshold are considered, while our objective is to identify the shortest route for efficient task delivery. In this simulation, we generate $64$ terrestrial E-MTs and $64$ UAV E-MTs, labeled from E-MT $1$ to E-MT $128$, and randomly select two integers to serve as the starting and ending points.

\begin{figure}
\vspace{-0.5cm}
    \centering
\includegraphics[width=0.65\linewidth]{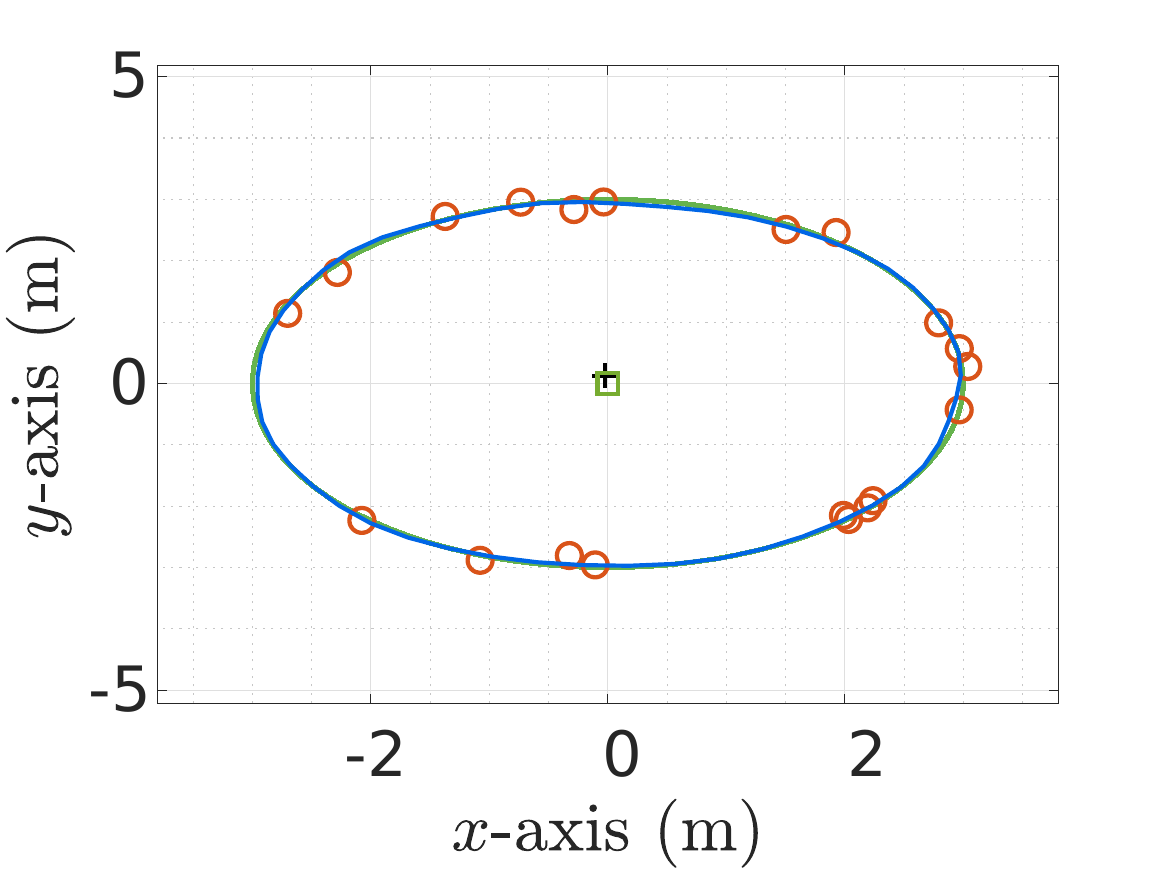}
    \caption{Elliptical-shaped extended-target localization using TA-based LAWN. The green rectangle is the true target location, while the black cross is its estimate. The red points indicate the estimated reflection points, the green curve represents the ground truth, while the blue curve shows the estimated target contour.}
    \label{fig-EOT}
    \vspace{-0.6cm}
\end{figure}

Figure \ref{fig:delay} compares the task transformation delays of three strategies, including: 1) TA-based Dijkstra method, 2) greedy-based locally optimal method, and 3) greedy-based locally reachable method. We see that the proposed scheme outperforms that of the greedy-based locally optimal and locally reachable methods. It is anticipated that the greedy-based locally optimal approach selects the nearest available node at each step. However, this method may fail to deliver the task from the starting point to the endpoint due to local minima or disconnected paths, resulting in a very low success rate of $1.22\%$, as shown in Fig. \ref{fig:delay}. To address this problem, the greedy-based reachable method can be applied. This approach incorporates re-selection mechanisms when the nearest option is unavailable, thereby increasing the success rate of task delivery, unless the destination is genuinely unreachable. In contrast, our TA-based method models the system as a sparse graph, where edge weights represent inter-node distances. This structure enables the direct application of the Dijkstra’s algorithm to search the globally optimal shortest path, ensuring reliable and efficient task transmission.

\vspace{-0.4cm}

\subsection{Extended-Target Localization}

Sensing is a core functionality of the LAWN and can be applied across diverse scenarios, including surveillance, target tracking, and environmental monitoring, as shown in Fig. \ref{fig:SystemModel}(e). In this subsection, we demonstrate the extended-target localization capability of the TA-based LAWN framework, aiming to recover both the shape and center position of the target. Specifically, $8$ E-MTs and the extended-target are deployed at the same altitude, while all other simulation parameters remain consistent with those in Section \ref{Subsection_Simulation_Setups}. During the extended-target sensing process, $50\%$ of E-MTs are designated as transmitters to emit probing signals, while the remaining functions are those of receivers to capture each signal from the target. Based on the received signals detected by the receivers, the angle-of-departure and time-of-arrival are obtained, which will then be used to estimate the reflection point positions. 
By collecting a sufficient number of reflection points, the Gaussian process \cite{EOT2015GP} is applied to recover the shape of the target, while the target position is acquired by averaging the collected reflection points.  

Figure \ref{fig-EOT} demonstrates that the shapes of an elliptical object can be accurately reconstructed in our framework. It is also seen that the estimated contours (blue curves) align closely with the ground truth (green curves), validating the effectiveness of the proposed scheme. It is anticipated that the positions of the E-MTs are flexibly coordinated by the D-MT, enabling both transmitters and receivers to steer signals toward different directions and capture reflections from multiple viewpoints. This multi-directional observation is essential for accurate extended-target sensing, as relying solely on reflections from one side of the target significantly degrades the accuracy of shape reconstruction. It should be noted that, in practical scenarios, the transmitter and receivers can be either UAVs or terrestrial BSs. Nevertheless, to fully exploit spatial diversity and enhance estimation accuracy, it is beneficial to assign both transmitting and receiving roles to UAVs. By optimizing the spatial placement of the participating UAVs via the D-MT, the LAWN system can attain high-precision estimation of both the shape and position of the extended-target.


\vspace{-0.35cm}

\section{Challenges and Possible Solutions}
\label{Section_Challenges}

Section \ref{Section_Case_Studies} shows that the TA-based framework effectively coordinates multiple MTs, enabling reliable execution of diverse network functions. However, fully unlocking the potential of TA-based LAWNs requires further investigation. In the following, we identify the potential challenges to real-world deployment and discuss their possible solutions.

\vspace{-0.4cm}

\subsection{Gap between Network Topology and Practical Network Implementations}

Although the proposed TA-based framework enables efficient multi-functional optimization by leveraging network topology, the abstraction process inevitably leads to partial information loss, as weak connections are deliberately excluded. Retaining more edges in the topological representation can preserve more network information, but it also significantly increases algorithmic complexity, posing a trade-off between graph density and computational efficiency. Moreover, certain studies in graph and topology theory treat all nodes and edges as equivalent, which may result in excessive and unnecessary resource consumption when applied to heterogeneous networks. As a result, the solution obtained from the TA-based formulation may differ from the global optimum of the original heterogeneous network.

\textbf{Possible Solution}: The performance of the developed scheme can be further improved by adopting more refined topological approximations and optimizing the underlying graph structure.
One possible solution is to introduce extra edges to mitigate information loss, but both nodes and edges can be associated with weights.
Another promising direction is to treat the topological structure itself as an optimization object, where only a subset of critical edges is selectively retained while others are disregarded. Such a topology refinement mechanism enables more accurate modeling without sacrificing scalability.

\vspace{-0.4cm}

\subsection{Cross-Component Coupling and Joint Optimization} 

In the LAWN, the system performance is affected by the mutual coupling between components. For example, a graph-theoretic topology optimized solely for throughput may favor dense inter-node connectivity. However, practical constraints, such as UAV flight endurance, collision avoidance, and sensing accuracy, may limit feasible node placement and link formation. By treating these constraints in isolation can lead to suboptimal or even infeasible network configurations, underscoring the need for a holistic design approach.

\textbf{Possible Solution}: An efficient optimization scheme that co-designs multiple network functions within a unified TA paradigm is required to address this problem. In addition, artificial intelligence (AI)-driven multi-objective optimization algorithms can further enhance coordinated and context-aware decision-making across system components.

\vspace{-0.4cm}

\subsection{Dynamic and Rapidly Evolving Network Topology}

The mobility of MTs leads to frequent topology changes, resulting in unstable links and fluctuating connectivity. Consequently, maintaining an accurate global or local view of the network topology incurs significant signaling overhead, particularly in dense aerial swarms. Moreover, LAWN wireless channels are highly sensitive to factors, such as altitude, urban blockages, terrain-induced reflections, and weather conditions, which contribute to non-stationary propagation characteristics. Fast fading and Doppler shifts further degrade link reliability, posing additional challenges in accurately forming and maintaining the network topology. To address these limitations, scalable TA routing and distributed network reconfiguration mechanisms are essential to ensure reliable connectivity with minimal control overhead.

\textbf{Possible Solution}: To handle rapid topology changes in LAWNs, the TA-based problem should transition from static link maintenance to mobility-aware strategies that dynamically adapt to node movement and evolving network conditions. By leveraging deep learning techniques, such as reinforcement learning and GNNs, the network can proactively anticipate link degradation through real-time channel estimation and prediction, enabling more resilient connectivity and efficient resource allocation in highly dynamic low-altitude environments.

\vspace{-0.3cm}

\subsection{Security, Privacy, and Trust in TA-Driven LAWN}

The mobility and openness of LAWNs increase vulnerability to security threats such as spoofing, jamming, and eavesdropping \cite{11315847}. The disclosure of topology and trajectory information can expose sensitive operational details, raising serious privacy concerns. In the proposed TA framework, adversaries may exploit dynamic topology changes to launch targeted attacks or infer mission-critical movement patterns. Moreover, command-and-control (C2) links require ultra-reliable and low-latency communication to ensure safe operation of low-altitude platforms; compromising these links can lead to mission failure or safety hazards, especially in congested or contested airspace.

\textbf{Possible Solution}: To strengthen the security level of TA-driven LAWNs, protection mechanisms should be tightly integrated with topology awareness. TA authentication and trust management can evaluate node reliability based on mobility patterns, link stability, and interaction history, enabling rapid detection and isolation of compromised nodes. In addition, abstraction of privacy-preserving topologies, such as anonymized routing, selective topology disclosure, and trajectory obfuscation, can reduce information leakage while preserving operational performance.

\vspace{-0.4cm}

\subsection{Standardization of LAWN}

The 3rd Generation Partnership Project (3GPP) has progressively strengthened support for low-altitude communication across recent releases. 3GPP Rel. 17 integrates UAV operations into the non-terrestrial network (NTN) framework, while 3GPP Rel. 18 further enhances UAV terminal capabilities, mobility management, and reliability mechanisms \cite{He20263GPP}. Although LAWN has not yet been formalized as a standalone architecture, these efforts provide partial foundations for aerial connectivity. Nevertheless, current standards do not fully address the unique requirements of LAWNs, where network performance depends on dynamic topology evolution and cross-layer adaptation. The absence of standardized topology representation, TA-based control signaling, and unified interfaces across heterogeneous aerial platforms remains a major barrier to interoperability, multi-vendor integration, and large-scale deployment.

\textbf{Possible Solution}: Future standardization efforts should incorporate explicit topology-aware primitives. By defining common interfaces for topology information exchange among aerial, terrestrial, and NTN layers can enable coordinated multi-tier network optimization. In addition, reference architectures and extensions of the TA-enabled protocol within 3GPP and related standard bodies will be critical to transform LAWN from an $ad$ $hoc$ aerial solution into a fully standardized and interoperable component of future 6G ecosystems.

\vspace{-0.35cm}

\section{Conclusion}
\label{Section_Conclusion}

To address the unique challenges of multi-functional and resource-constrained LAWN, this article devised a TA coordination framework within a heterogeneous LAWN. By representing the LAWN as a graph, the proposed TA approach enables efficient task allocation and multi-functional cooperation among heterogeneous devices without incurring the prohibitive overhead of constant network-wide information exchange. Simulation results validated the effectiveness of our framework, demonstrating notable improvements in efficiency and performance compared to baseline schemes. Future work will focus on enhancing the representation of LAWN topology and integrating machine learning techniques to enable cross-layer resource optimization and adaptive security, thereby extending its applicability to emerging services, such as digital twins and extended reality.
\vspace{-0.5cm}
\bibliographystyle{ieeetr}
\bibliography{ref}
\vspace{-1cm}
\begin{IEEEbiographynophoto}
{Jiajun He} is a Research Fellow with Queen’s University Belfast (QUB), United Kingdom.
\end{IEEEbiographynophoto}
\vspace{-1.2cm}
\begin{IEEEbiographynophoto}
{Han Yu} is a Research Fellow at the Faculty of Electrical Engineering and Computer Science, Technical University of Berlin, Germany.
\end{IEEEbiographynophoto}
\vspace{-1.2cm}
\begin{IEEEbiographynophoto}
{Yiran Guo} is a Ph.D. student in Beijing Jiaotong University, China.
\end{IEEEbiographynophoto}
\vspace{-1.2cm}
\begin{IEEEbiographynophoto}
{Xinping Yi} is a Professor with the School of Information Science and Engineering, Southeast University, Nanjing, China
\end{IEEEbiographynophoto}
\vspace{-1.2cm}
\begin{IEEEbiographynophoto}
{Fan Liu} is a Professor with the School of Information Science and Engineering, Southeast University, Nanjing, China.
\end{IEEEbiographynophoto}
\vspace{-1.2cm}
\begin{IEEEbiographynophoto}
{Hing Cheung So} is a Professor with Electrical Engineering, City University of Hong Kong, Hong Kong, China.
\end{IEEEbiographynophoto}
\vspace{-1.2cm}
\begin{IEEEbiographynophoto}
{Hien Quoc Ngo} is a Professor at Queen’s University Belfast (QUB), United Kingdom.
\end{IEEEbiographynophoto}
\vspace{-1.2cm}
\begin{IEEEbiographynophoto}
{Michail Matthaiou} is a Professor at Queen’s University Belfast (QUB), United Kingdom.
\end{IEEEbiographynophoto}
\vspace{-1.2cm}
\begin{IEEEbiographynophoto}
{Giuseppe Caire} is a Professor at the Faculty of Electrical Engineering and Computer Science, Technical University of Berlin, Germany.
\end{IEEEbiographynophoto}



















\end{document}

%% file: my_macros.tex
\usepackage{amsthm,amsmath,amssymb}
\usepackage{eucal}
\usepackage{xifthen}
\usepackage{mathtools}
\usepackage{enumerate}
\usepackage{microtype}
\usepackage{xspace}
\usepackage{bm}
\usepackage{cite}
\usepackage{fancyhdr}
\usepackage{lastpage}
\usepackage[small]{caption}
\usepackage{xcolor}
\allowdisplaybreaks

\long\def\comment#1{}

\newfont{\bbb}{msbm10 scaled 700}

\newfont{\bb}{msbm10 scaled 1100}

\renewcommand{\Re}[1][]{\ifthenelse{\isempty{#1}}{\operatorname{Re}}{\operatorname{Re}\left(#1\right)}}
\renewcommand{\Im}[1][]{\ifthenelse{\isempty{#1}}{\operatorname{Im}}{\operatorname{Im}\left(#1\right)}}

\newcommand{\CN}[1][]{\ifthenelse{\isempty{#1}}{\mathcal{N}_{\mathbb{C}}}{\mathcal{N}_{\mathbb{C}}\left(#1\right)}}
\renewcommand{\P}[1][]{\ifthenelse{\isempty{#1}}{\mathbb{P}}{\mathbb{P}\left(#1\right)}}
\newcommand{\E}[1][]{\ifthenelse{\isempty{#1}}{\mathbb{E}}{\mathbb{E}\left(#1\right)}}
\renewcommand{\det}[1][]{\ifthenelse{\isempty{#1}}{\mathrm{det}}{\mathrm{det}\left(#1\right)}}
\newcommand{\trace}[1][]{\ifthenelse{\isempty{#1}}{\mathrm{tr}}{\mathrm{tr}\left(#1\right)}}
\newcommand{\rank}[1][]{\ifthenelse{\isempty{#1}}{\mathrm{rank}}{\mathrm{rank}\left(#1\right)}}
\newcommand{\diag}[1][]{\ifthenelse{\isempty{#1}}{\mathrm{diag}}{\mathrm{diag}\left(#1\right)}}

\renewcommand{\Re}{{\rm Re}}
\renewcommand{\Im}{{\rm Im}}

\allowdisplaybreaks
\DeclareMathAlphabet{\mathcal}{OMS}{cmsy}{m}{n}

\usepackage{hyperref}
\hypersetup{
    bookmarks=true,         
    unicode=false,          
    pdftoolbar=true,        
    pdfmenubar=true,        
    pdffitwindow=false,     
    pdfstartview={FitH},    
    pdfnewwindow=true,      
    colorlinks=true,       
    linkcolor=red,          
    citecolor=green,        
    filecolor=magenta,      
    urlcolor=cyan           
}